\documentclass[12pt,preprint]{aastex}

\newcommand{\msun}{M_{\odot}}
\newcommand{\etal}{{\it et~al.}}

\shorttitle{X-ray Spectra of SMC X-1}
\shortauthors{Vrtilek et al.}

\begin{document}


\title{Photoionized lines in X-ray Spectra of SMC X-1} 


\author{S.D. Vrtilek, and J.C. Raymond}
\affil{
Harvard-Smithsonian Center for Astrophysics}
\email{svrtilek@cfa.harvard.edu}

\author{B. Boroson}
\affil{Bowling Green State University}

\and

\author{R. McCray}
\affil{JILA}


\begin{abstract}
We present a detailed spectral analysis of Chandra/ACIS-S CC mode
observations of the massive X-ray binary system SMC X-1.  The system 
was observed during both the high and low X-ray states of the roughly 
60-day superorbital period.  The continuum spectra during both states 
are well represented by a power law with photon index $\alpha$=0.9 and 
a blackbody of kT~=~0.15keV.  The high state spectra are dominated by  
the continuum and independent of orbital phase whereas the low state 
spectra show a strong orbital dependence as well as line emission from 
O, Ne, Mg, Fe, and Si.  This is consistent with the states attributed 
to disk precession: during the high state X-ray emission is dominated 
by the compact source which is abrubtly eclipsed and during the low 
state the compact object is hidden by the disk and a larger, less 
luminous scattering region is responsible for the X-ray emission.  A 
prominent Ne IX feature places a stringent limit (Log $\xi$ = 2.0-2.5)
on the ionization parameter which constrains the wind dynamics of the  
system. The Fe line fluxes are related linearly to the blackbody 
fluxes indicating that both originate in the same region or are excited 
by the same mechanism.  There is evidence for structure in the Fe-line 
that cannot be fully resolved by the current observations.  The pulse 
period measured during our observations, 0.7057147$\pm$0.00000027s 
shows that the uninterrupted spin-up trend of SMC X-1 continues.
We discuss the implications of our results for models of SMC X-1.

\end{abstract}


\keywords{accretion, accretion disks --- binaries: close --- pulsars: individual (SMC X-1) --- --- X-rays: stars}


\section{Introduction}

The massive X-ray binary system SMC X-1/SK160
is one of only two sources known to show both pulses and bursts;
\citet{li97} suggest that this is due to
the magnetic moment of SMC~X-1, $\sim$10$^{29}$G cm$^3$, which
is an order of magnitude
lower than those of typical X-ray pulsars.
SMC X-1 is the only X-ray pulsar for which no
spin-down episodes have been observed \citep{kah99}. If
the magnetic moment of SMC X-1 is as low as Li and van den Heuvel suggest,
then
the spin-up trend can be explained by classical accretion torque models
\citep{kah99}.
In addition to showing X-ray pulses with a 0.71s period,
SMC X-1 is eclipsed by its
B0 supergiant companion for $\sim$ 15 hours during every 3.89 day
binary orbit.
The uneclipsed X-ray flux exhibits aperiodic variabilities on
timescales from milliseconds
to months.  These include quasi-periodic oscillation at 0.06 Hz
\citep{woj98},
X-ray bursts \citep{ang91}, and a
longterm ``superorbital'' period of 50-60 d, reported by
\citet{gru84} and
confirmed by RXTE
\citep{lev96} and several other X-ray observatories
\citep{woj98}.
For SMC X-1 we assume that the long-term period is due to a warping
and precession of the disk induced by the effects of uneven X-ray
irradiation on the disk, as is the case for Her X-1 and LMC X-4
\citep{che95,vrt97}
and as is consistent with
requiring a classical accretion torque model for SMC X-1.

The ultraviolet spectra of SMC X-1 are similar to those of
the massive X-ray binary system LMC X-4 and its
15 $\msun$ O star companion; both show ultraviolet P-Cygni lines that
vary dramatically with orbital phase, with strong broad absorption near
X-ray eclipse and narrow absorption when the X-ray source is in the
line-of-sight \citep{vrt01}.  The latter have been interpreted as a
result of X-ray photoionization
of the stellar wind; when the neutron star is in front of the normal
star, the wind absorption disappears and mainly the photospheric
absorption lines are visible \citep{vrt97,bor99}.
Detailed modelling of the P-Cygni line profiles of LMC X-4
provides measurements
of the stellar wind parameters and indicates the presence of
inhomogeneities in the wind \citep{bor99}.

Here we report on our analysis of the pulse
averaged X-ray data.
In Section 2 we describe the observations, in Section 3 we present the
results from our spectrum analysis, and in Section 4
we consider the implications of our results and future work.

\section{Observations}

Eight observations of SMC X-1 were made with the ACIS-S on Chandra
from 2000 October to 2001 April; these were coordinated
with 10 observations
made by HST/STIS.  Preliminary results of the multiwavelength campaign
were presented in \citet{vrt01}.
Here we present a detailed spectral analysis
of the X-ray observations as listed in Table 1.
The orbital phases listed in Table 1 were calculated using
the orbital period and ephemeris given by \citet{woj98} with
T$_0$ updated for N=2310 (where N is the number of orbits since the
ephemeris of Wojdowski~\etal)
relative to their ephemeris:\\

T$_0$(JD) = 2451827.50374 + 3.8922909N - 6.953E-8N$^2$

\vskip 0.1in

Since SMC X-1 is a bright source the CCD detectors were subject to
``pileup": this occurs when two or more photons are incident on the
same pixel during the CCD time-resolution element or ``frame-time.''
In order to
mitigate this effect we used ACIS-S3 in the continuous clocking (CC) mode.
In CC mode the
frame-time is .00855s.
This proved sufficient during
the X-ray low states and during eclipse but significant pileup
fractions occurred during the X-ray high state.
The high-state out-of-eclipse countrates (40-54 cts/sec)
translate to 0.4-0.5 cts/frame
corresponding to pileup fractions of 16-18\%.
During eclipse and the low-state we had less than
.03 cts/frame and pileup is
less than 1\%.
The source counts were extracted from a rectangular area 4 by 1 pixels
from the CC image and an equivalent area spaced 100 pixels away was used
for the background.  The background was negligible during
the high state and
less than 3\% of the source counts during the low-state.

The data were processed using the CIAO version 3.0.3. 
data analysis package. 
The current analysis includes a correction for spacecraft
dither that was not available
when the preliminary results were published; in addition, observations
affected
by ACIS pileup
were corrected.
We deviated from the
standard ACIS processing since the pipeline program 
``acis\_detect\_afterglow''
misidentifies up to 20\% of actual events
as cosmic ray afterglow events.
The source photons are rejected in systematic, non-uniform ways.
In order to correct for pileup we
start with data files before they
undergo standard processing (level 1 files).
We then use MKARF to construct an ARF appropriate for our
observation. MKARF takes
observation-specific dither and bad pixels into account, and
determines
fractional exposure time lost to dithering
over bad pixels.




\section{Spectral Analysis}

Analysis of products produced with CIAO tools was performed with the
Interactive Spectral Interpretation System (ISIS); \citep{hou00}
and the pileup
kernel developed by \citep{dav01a,dav01b} and incorporated into ISIS by
\citet{hou02} was utilized.

The pulse-averaged (0.1-200keV) X-ray spectrum of SMC X-1 during the
high-state of the superorbital cycle has been described as a power-law
($\alpha=0.9$) with a high-energy
cutoff (5.6keV with a 15~keV folding energy), a thermal component
(blackbody with kT~=~0.15keV) at low energies, and Fe~K-shell emission
\citep{woo95}.
During X-ray eclipse \citet{woj00} fit
ASCA data to a power law with photon index ($\alpha$=0.94)
and a high-energy cutoff at 10 keV with a 15 keV folding energy.
\citet{nai04} fit BeppoSax observations of SMC X-1 with a power-law
of photon index ($\alpha$=0.8-0.9), blackbody with kT=0.16-0.19, and
a high energy cutoff at 6.3 keV with a 11~keV folding energy.
In all cases we include 
the equivalent neutral column density (N$_H$) using Wisconsin \citep{mor83}
cross-sections:
N(E) = exp$^{-N_H\sigma(E)}$
where $\sigma(E)$ is the photo-electric cross-section (excluding Thomson 
scattering) with \citet{and82}
relative 
abundances. 

These models are similar to those used to fit the closely related
systems LMC X-4 and Her X-1 \citep{hic04,zan04} 
with the exception that neither
LMC X-4 nor Her X-1 require high energy cutoffs withing the Chandra
energy range.
Since the cutoff energy and e-folding energy invoked for SMC X-1
differ for different instruments as our first attempt in fitting
the continuum for SMC X-1 we use a simple power law plus a blackbody.

\subsection{Fits to High-state Spectra}

Because of the limited energy range of the ACIS-S
in fitting the high state spectra we
kept the power law photon index at the value ($\alpha$ = 0.9) found by 
several observers using different instruments 
\citep{woo95,woj98,nai04}.
We note that when the photon index is allowed to vary it
converges to values between 0.8 to 1.0.
We used the energy range 0.6-9.8 keV for fits but excluded the region 
1.7-2.8 keV
to take into consideration
instrumental effects:
the reflectivity of the HRMA is presently underestimated by about
10-15\% around the IR-M edge at 2.1 keV
(a likely source of the
problem is a 7 $\AA$ layer of carbon on the mirrors which increases
the reflectivity near edges; L. David personal communication).
This is particularly troublesome for high count rate data.
The blackbody temperature was allowed to vary and converged to the
value found by the earlier observers.
This simple continuum model (hereafter Model 1) provided reasonable fits to
 the
high-state data as shown in Figure 1 for the parameters listed
in Table 2a.
The spectra are remarkably independent of orbital phase during the 
high-state.
This suggests that during the high state the X-ray emission is dominated
by the NS and it gets eclipsed rapidly by the companion.
During the low state the NS is not directly in our line of sight
and the diffuse material that then dominates is eclipsed gradually.

Although the simple power-law plus blackbody provided good fits to the
high-state continuum spectra, since previous observers had included 
high-energy
cut-offs we also fit our data to models that included this additional
constraint (Model 2).   We found that the cutoff energy and e-folding 
parameters
could not be constrained by our data and converged to different values
depending on the initial guess.
Table 2b shows the results when the powerlaw photon index is allowed to
vary; holding the photon index at 0.9 did not result in better constraints
on the high energy cutoff and e-folding values.

Finally,
we used a model in which soft photons are Comptonized in a hot plasma
as generalized by \citet{tit94}.
This model (Model 3: ComptT) provided fits that were comparable to the 
simple power-law plus
blackbody model in the sense that the reduced chisq converged to 1.
However, the individual parameters of the model were not well constrained 
by the data and converged to different values depending on the initial
guess.

We conclude that a blackbody of around 0.15~keV provides an excellent fit
to the low-energy flux from SMC X-1 but that the current data are not
of sufficient quality and energy range to uniquely constrain the 
high-energy flux.

\subsection{Fits to Low-state Spectra}

Since the eclipse and low-state spectra show evidence of spectral
lines we augmented the continuum models with bright recombination
lines determined using the XSTAR code \citep{kal99}.
We used as input to XSTAR our best-fit continuum models with
different values of the ionization parameter $\xi$=L$_x/(nr^2)$
\citep{tar69};
where L$_x$ is the X-ray
luminosity, $n$ is the proton number density, and r is the
distance from the ionizing source).  We used a metal abundance
of 1/5th solar as expected for the SMC \citep{woj00}.
Our observations are consistent with emission from a photoionized
plasma when the ionization parameter is in the range
2.0~$\ge$~log~$\xi$~$\le$ 2.5.
We further restricted ourselves to lines that had been
reported in the literature (using higher energy resolution data than the 
ACIS-S) for HMXB systems similar to SMC X-1.
Since both models \citep{jim01} and observations of
several similar systems \citep{sak99,pae00}
indicate that radiative recombination continua (RRC:
an unambiguous indicator of excitation by
recombination in X-ray-photoionized gas) 
should be
present at strengths up to half the intensities of the Lyman
lines we have included reported RRC in our fits.
Table 3 gives a list of lines and RRC that were searched for
to fit our observations.
In each case we added lines using the
F test criterion (F$_x >$ F) for inclusion of additional terms \citep{bev69}:

$$F = \chi_{\nu}^2(n-1)/\chi_{\nu}^2(n)$$

$$F_x = [\chi^2(n-1) - \chi^2 (n)]/[\chi^2(n)/(N-n-1)]$$

The observed countrates varied by a factor of 60 during our observations.
In order to ensure sufficient counts per bin for spectral fitting low count
spectra were
binned before fitting:
typically for the low state N was 125.
Since we are dealing with relatively low resolution CCD data
the line
widths were constrained at 50 eV; hence only the line strength was
a free variable.
Some of the lines listed in Table 3 are blended in ACIS (for example we can
not
distinguish Ne~IX [0.91 keV] from O~VIII RRC [0.89 keV] or 
O~VIII~Ly$_{\beta}$ 
[0.77 keV] from O~VII RRC [0.74 keV]).
However, the O VIII~Ly$_{\alpha}$ should be comparable to the RRC, and it
is weak.
At orbital phase 0.2
the feature at 2.4 keV may be identified as Si~XIII RRC but
Si~XIII Ly$_{\alpha}$ (2.0 keV)
is below our detection limit.

For our first attempt
we used the simple power-law plus blackbody
used for the High State augmented with the lines listed in Table 3 
(Model 1+L). While this model provided
good fits to the data (Table 4a) we noticed that the residuals show an
excess near 6 keV that could not be modeled with narrow line
emission.
ASCA observations of SMC X-1 showed a broad ($\sigma$ = 0.85 keV) feature 
at 6 keV
that could be attributed to Compton scattering of Fe K \citep{sta97}.
Since
excess emission near this energy has been modeled with a skewed
broad line (associating
the broad line with the compact object) for several HMXBs
\citep{mil02a,mil02b,mil04a,mil04b} we tried adding
this feature to our models.
We found that the fits were significantly improved with the addition of a 
skewed broad feature.
In Table 4a we list the reduced $\chi^2$ when a broad ($\sigma$ = 800 eV)
feature at 5.8 keV
is included in the fits.
Figures 2-4 show the low state data and our best fits with (Model 1+L+B) 
and without
inclusion of a broad Fe line.

For completeness we fit the low-state data to all the models
used for the high-state data.
Again, for Model 2 the high-energy and e-folding values could not be 
constrained.
\citet{sta97} also found that the ASCA data showed poor fits when 
using a cutoff
powerlaw rather than a simple powerlaw plus broad gaussian.
The results for Model 3 as listed in Table 4b and
shown in figures 2-4 provided formally good fits to the data {\it without}
requiring broad Fe emission {\it but} again the individual parameters are 
not well constrained.

\subsection{Fits to Eclipse data}

Since correction of spacecraft dither effects had not been implemented
for CC mode data
during our initial analysis of the eclipse data presented in
\citet{vrt01} we
have reanalyzed them.  The best fit parameters are listed in
Table 5 and plotted in Figure 5.
The main effect of the correction is a reduction of background counts
which resulted in a better determination of the continuum and 
correspondingly
better confidence in the line detection.
No broad Fe features were required during eclipse
as is consistent with associating the broad feature (if it is real) 
with the neutron star.

\subsection{The Fe line}

For Model 1 both the Fe K$_{\alpha}$ and broad Fe line fluxes are
positively correlated with the
black body flux indicating either that they arise in the same region or
are excited by the same process (Figure 6).  If we interpret the
blackbody radiation as hard X-rays reprocessed in the disk following
\citet{hic04}, the emission
radius ($R_{bb}^2$ = L$_x \div [4\pi \sigma T_{bb}^4]$) 
is a few times
10$^8$~cm, corresponding to the inner edge of the accretion disk.
This interpretation considers a thick, partially spherical shell centered
on the neutron star and subtending a solid angle, $\Omega$, at the X-ray source
and differs from some estimates for R$_{bb}$, which assume the soft
luminosity is radiated from a full sphere at R$_{bb}$ (see \citet{hic04} Figure 9).
This is consistent with the pulse profile analysis of these data by
\citet{nei04}: they find that the soft pulses and the blackbody 
component both originate in the inner edge of the accretion disk.
For Model 3 there is no particular correlation between the Fe-line and
either of the continuum fluxes).

In Figure 7 we show closeups of the Fe-line regions of the low state
spectra showing the narrow
and broad components.  For Model 1
the profiles are similar to Fe line profiles seen
in AGN and Galactic
black hole candidates.
since we had restrained our fits to photon index $\alpha$=0.9 whereas
allowed range during the high state is 0.8=1.0 we computed our values
for the extremes of photon index as listed in Table 4b and plotted
in Figure 8.
The fact that the ComptT models do {\it not} require broad Fe lines
suggests that ACIS-S cannot resove
the Fe-line profile.  The low count statistics and limited energy range
(which makes continuum determination difficult) are also factors requiring
consideration.
We therefore caution the interpretation of observations taken with 
instruments 
at the resolution of the ACIS-S or poorer that claim to detect broad
Fe lines unless the continuum is well-constrained by other means.
The factor of 10 improvement in energy resolution and
collecting power of ASTRO-E2 over the Chandra HETG near 6 keV may help 
resolve this issue.

Zooming in on the Fe line region for the eclipse observations (Figure 9) 
shows
both hydrogenic and helium-like Fe for the eclipse during the high state
whereas only Fe K$_{\alpha}$ is seen during the low-state eclipse.
There is no hint of a broad Fe feature.

\section{Discussion and Conclusions}
The gross structure of accretion disks and mass transfer has
been explained as a
problem governed by classical hydrodynamics with
boundary conditions set by the masses and separation of the two stars
\citep{sha73,lub75}.
However complications arise because X-ray heating is important even in
the presence of strong viscous heating in the disk interior
\citep{dub99} and instabilities driven by the radiation
play a significant role on the dynamics 
\citep{pri96,mal96}.  Important physical
processes in accretion disks, such as the viscous
transport of angular momentum,
the generation of magnetic fields, the nature of disk coronae and winds
\citep{beg83,bms83}, and turbulence
within the disk and in the interaction with the gas stream from the
companion \citep{arm98} have been characterized
or parameterized,
but our understanding is still in an exploratory stage.

In addition for high-mass X-ray binaries (HBXBs) such as SMC X-1
the stellar wind plays an active role as a partial source of
fuel for the X-ray luminosity; the wind also plays a passive
role because the X-ray spectrum is modified as it passes through
the wind \citep{blo95}.

One path toward obtaining clarifying information on these open issues is
the study of discrete spectral features. 
 The CCD on Chandra provides
sufficient resolution to identify some of these features,
although we note that higher energy resolution observations are
necessary to exploit fully the information contained in them.
For example, the detection of more than two dozen recombination emission lines 
in Chandra HETG spectra of Her X-1 allowed \citet{jim04} to measure the density, temperature,
spatial distribution, elemental composition, and kinematics of the plasma; the absence
of detected Doppler broadening in the lines allowed them to constrain the dynamics of the disk
atmosphere and the geometry of the disk.  
XMM RGS observations of recombination lines in the eclipse spectrum of high-mass X-ray binary,
4U1700-37, allowed \citet{van04} to conclude that the formation region extends beyond
the size of the O supergiant. 
Observations of SMC X-1 with either the HETG on Chandra or the XRS on ASTRO-E2 (which provides
factor-of-10 improvements in energy resolution and
collecting power over the HETG near 6 keV) promise to extend this work
in significant ways.  

Our continuum fits during the high and low X-ray states are 
consistent with precession of an accretion disk in SMC X-1.
During the high state the continuum displays little or no orbital 
variation out of eclipse:  this is consistent with the central
source being directly in our line-of-sight. The X-ray flux is reduced
abruptly during eclipse.  During the low-state the central
source is blocked by the precessing accretion disk and we
see emission from a larger scattering area with a radius of a few times
10$^8$ cm corresponding to the inner edge of the accretion disk;
this larger area is gradually covered leading to the observed
orbital variation.  The X-ray flux during the low-state eclipse
is roughly half that during the high-state eclipse because half
the larger scattering area is obscured by the disk blocking
the central source. 

We have detected lines due to O, Ne, Mg, Fe, and Si as
well as a possible RRC feature from Si in the low-state X-ray
spectra of SMC X-1 using the Chandra ACIS-S in CC mode.
A stringent limit on the ionization parameter
is placed by the prominent Ne IX feature 
(Log $\xi$~=~2.0-2.5;
see Fig. 5 in \citet{woj01}).  Because ASCA's energy resolution was
not sufficient to detect these lines \citet{woj01} had concluded that
the ionization parameter was either greater than 3 or less than 1 
requiring the presence of circumstellar gases with both high and low ionization.
Our result solves this problem and is consistent with the 
density distribution derived by \citet{blo95}
using three-dimensional hydrodynamic simulations of the SMC X-1
system.  This value of the ionization parameter is also consistent
with the Hatchett-McCray effect seen in ultraviolet lines observed 
during our simultaneous observations
with HST/STIS \citep{vrt01} as well as recent observations of the
OVI 1032 line with FUSE showing a similar Hatchett-McCray effect
\citep{ipi04}. 

\citet{jim01} have modeled X-ray line emission from
accreting disk atmospheres.  They find a clear difference between
evaporating or condensing disk atmospheres that should be detectable for
either high-inclination systems or for systems where the central source 
is obscured but
the extended disk emission is still visible: for example the
O~VIII Ly$_{\alpha}$ to O~VII He$\alpha$ flux ratio in a condensing
disk model is five times that of the evaporating disk model; and
O~VII RRC for the evaporating disk model has two times the flux
predicted for the condensing disk model.
At orbital phase 0.6 during the low state (which corresponds to a time
when the central source is obscured) our O~VIII Ly$_{\alpha}$ to
O~VII He$\alpha$ flux ratio
is consistent within errors with that expected for an evaporating disk 
model.  We do not have the sensitivity to detect
the O~VII RRC feature at 0.74 keV.

During orbital phase 0.0 in the high-state we find evidence
for both hydrogenic and helium-like Fe in addition to the resonance line
with relative ratios similar to those reported for Cen X-3 by Wojdowski
\etal~(2003).
However, the relatively poor resolution of the CCD does not allow us to
determine line ratios accurately.  The fact that these 
features
are not detected during low state eclipse is consistent
with the fact that we see considerably more of the scattering
material during the high state.  The low state eclipse is effectively 
a fraction of the high state spectrum because the scattering region 
around the 
B star sees more of the disk.  Grating observations are needed in order
to test if the resonant X-ray line scattering model proposed for Cen X-3 by
\citet{woj03} is applicable to SMC X-1.

Finally, we have measured the pulse period of SMC X-1 during
the epoch of our observations and find that it continues to
show steady spin-up at an average rate of -1.2$\times~10^{11}$s~s$^{-1}$
(Figure 10).
SMC X-1 thus continues to be the only X-ray pulsar for which no
spin-down episodes has been observed.
Pulse profile analysis of the high state data show a hard pulse
profile that is consistently double-peaked with a soft pulse
profile that varies in shape between single and double peak
\citep{nei04}.  Although the data cover a small fraction
of the superorbital cycle these profiles are consistent with the
a model in which the soft pulses occur by reflection of the
hard pulses at the inner edge of the accretion disk with
different segments of the inner edge in the observer's line of view
as a function of superorbital phase.  This picture is consistent with
the soft flux (blackbody) coming from the inner edge of the
accretion disk as well as the Fe line.







\acknowledgments
We are grateful
to the Chandra and
HST mission planning teams for a superb job coordinating the two satellites
for this project.  SDV would like to acknowledge partial 
support from NASA grants
NAG5-6711, G0-08566.01-A through STScI, and G01-2028X through CXC.

\clearpage


\begin{figure}
\plotone{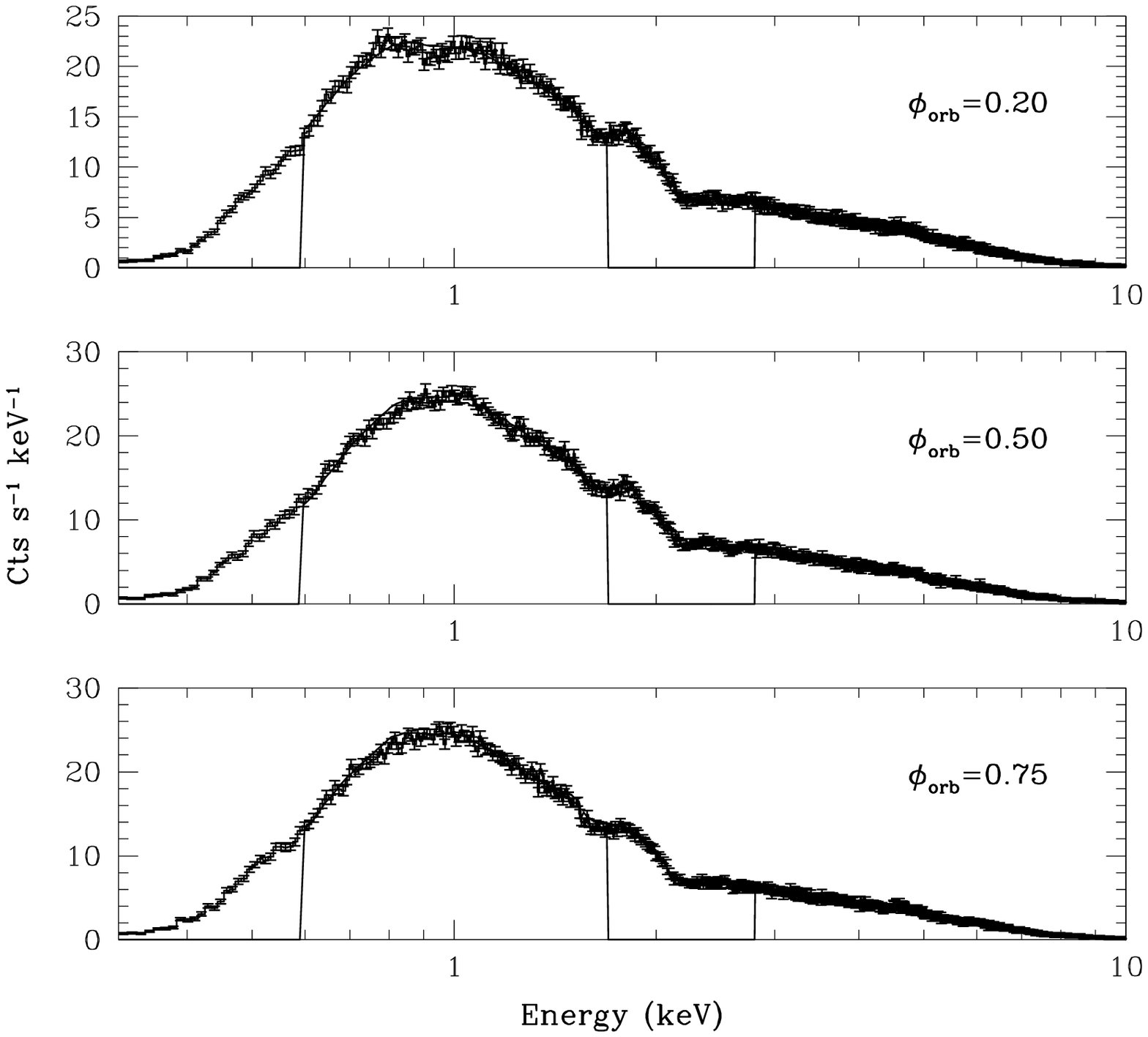}
\caption{
Chandra ACIS-S spectra obtained during X-ray high state:
fits to a simple continuum model of power law plus blackbody and
neutral absorption.
The histograms represent the data and the smooth lines the best fit
continuum models as listed in Table 3.
The spectra are fitted from 0.6-1.7 and 2.8-9.8 keV only.
}
\end{figure}

\clearpage 

\begin{figure}
\plotone{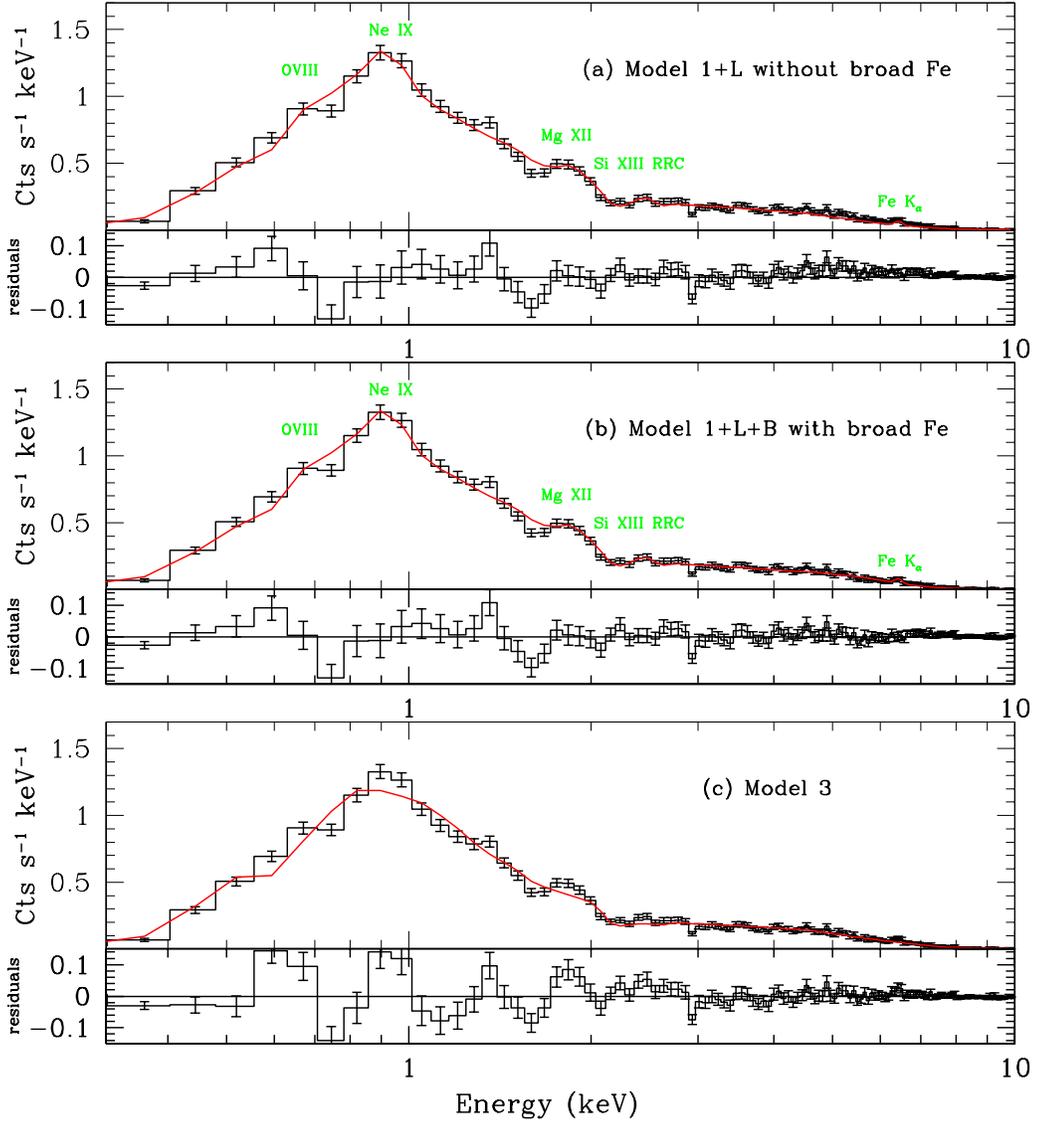}
\caption{Low-state orbital phase 0.2.
The histograms represent the data and the smooth lines the best fit
models as listed in Tables 4 and 5.
(a) (Model 1+L) Powerlaw plus blackbody plus lines excluding broad Fe.
Residuals show systematic excess near Fe.
(b) (Model 1+L+B) Powerlaw plus blackbody plus lines including broad Fe.
(c) (Model 3+L) ComptT plus lines.
The residuals are in the same units as the data.}
\end{figure}

\clearpage
\begin{figure}
\plotone{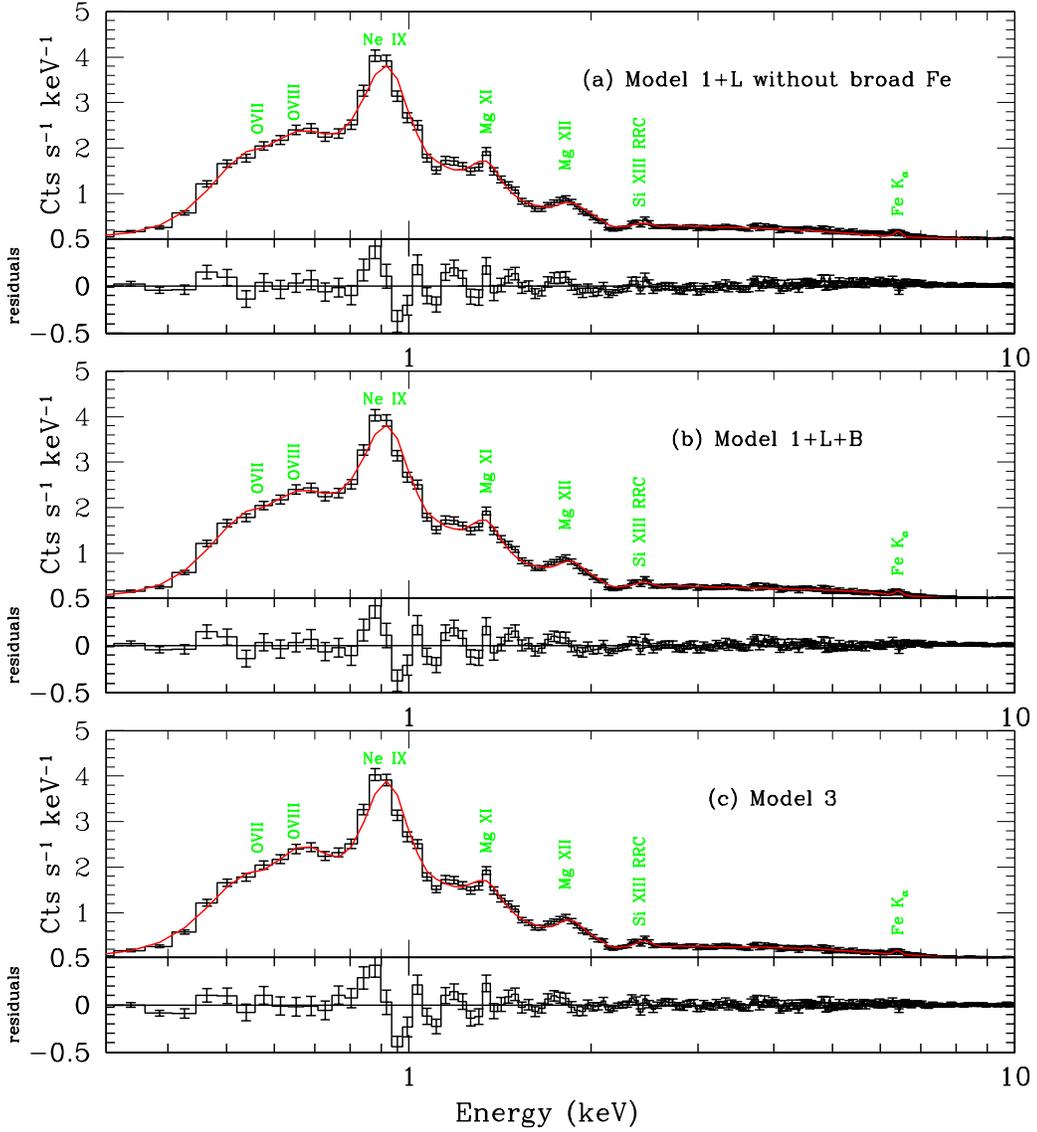}
\caption{
Low-state orbital phase 0.6.
The histograms represent the data and the smooth lines the best fit
models as listed in Tables 4 and 5.
(a) Model 1+L excluding broad Fe.
Residuals show systematic excess near Fe.
(b)   Model 1+L+B including broad Fe.
(c) Model 3+L. 
The residuals are in the same units as the data.
}
\end{figure}

\clearpage

\begin{figure}
\plotone{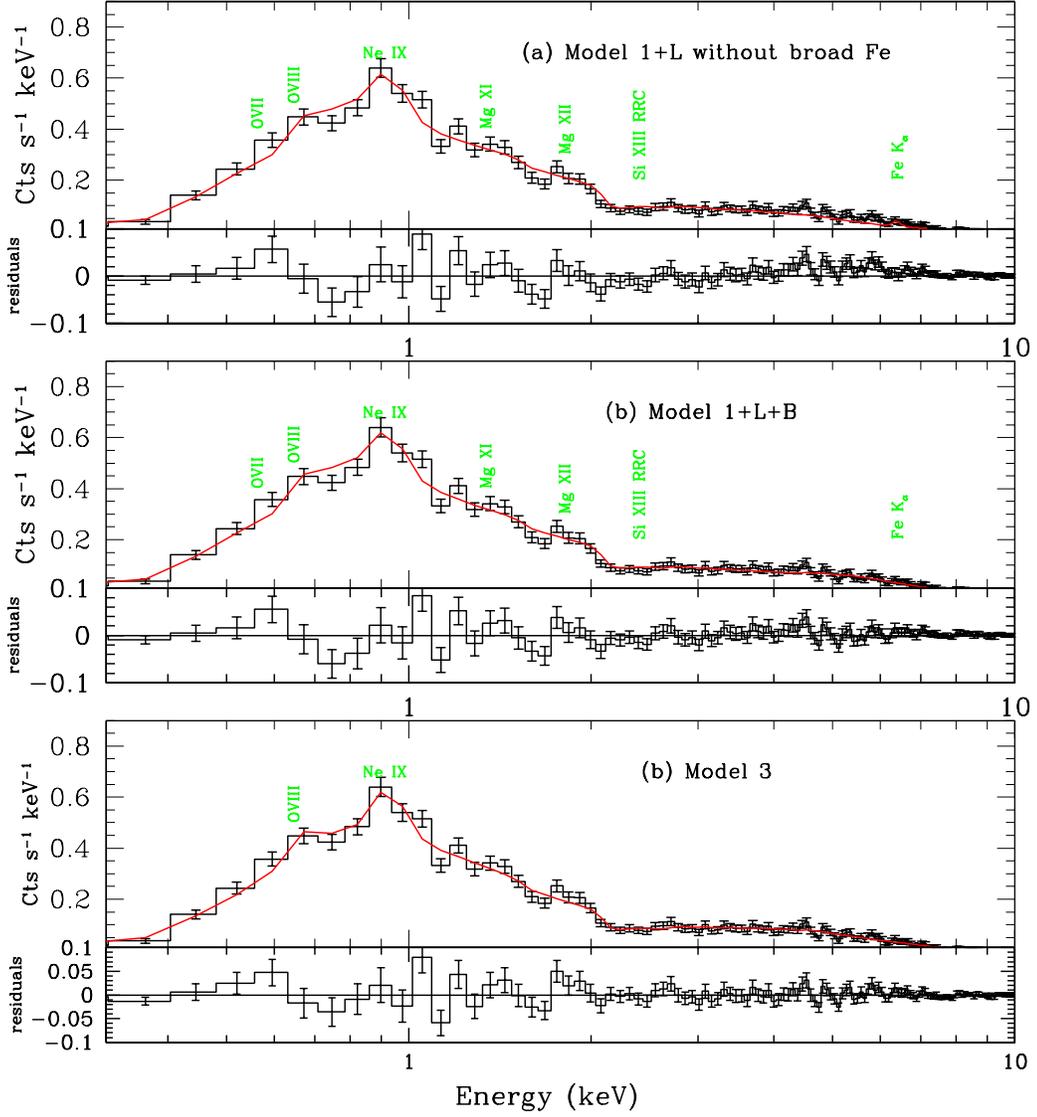}
\caption{Low-state orbital phase 0.8.
The histograms represent the data and the smooth lines the best fit
models as listed in Tables 4 and 5.
(a) Model 1+L excluding broad Fe.
Residuals show a narrow feature at 5.8 keV.
(b)   Model 1+L+B including broad Fe.
(c) Model 3+L. 
The residuals are in the same units as the data.
}
\end{figure}

\clearpage

\begin{figure}
\plotone{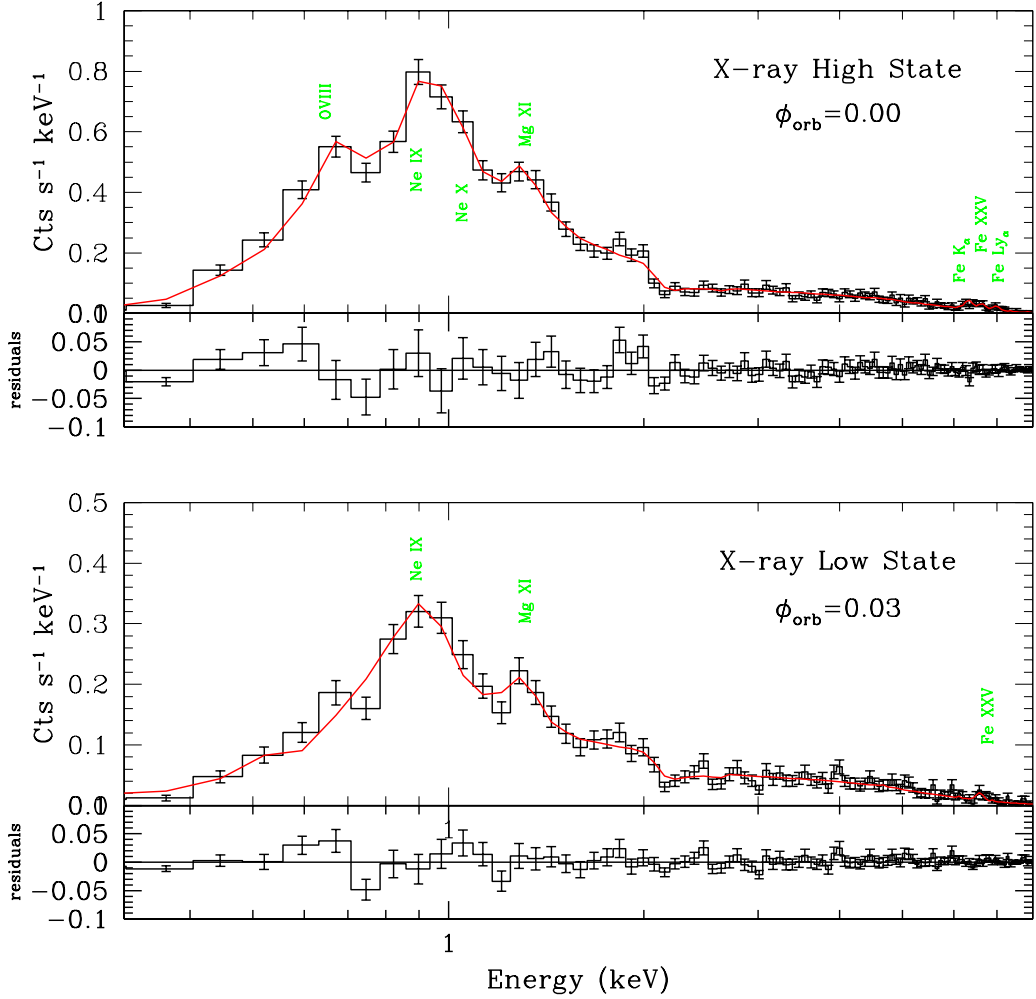}
\caption{
Chandra ACIS-S spectra obtained during X-ray eclipse:
X-ray state and orbital phase are indicated for each spectrum.
The histograms represent the data and the smooth lines the best fit
models as listed in Table 4.
The residuals are in the same units as the data.}
\end{figure}

\begin{figure}
\plotone{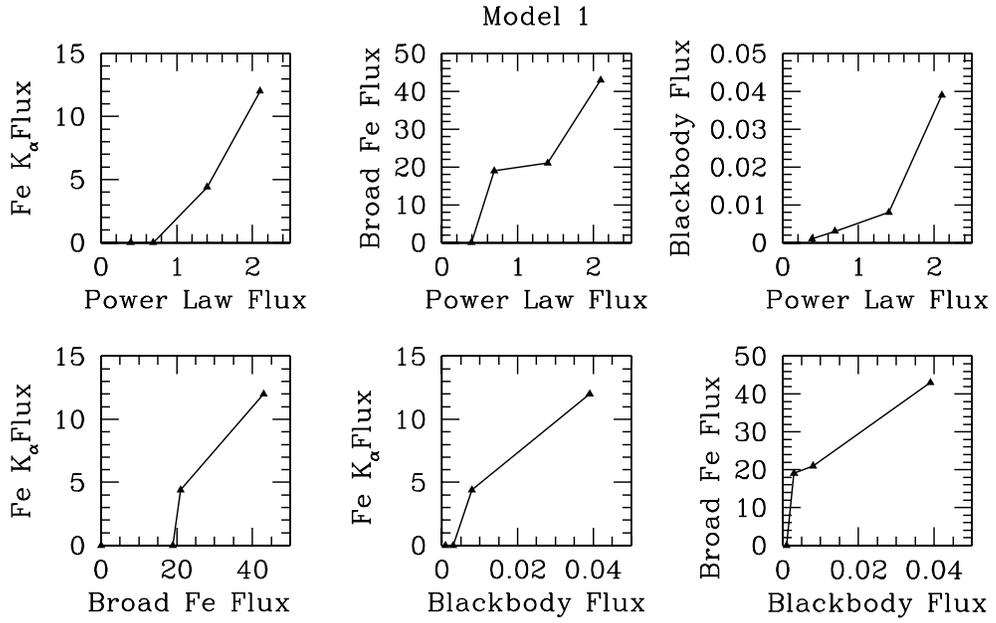}
\caption{\it
(a) Power law flux versus Fe K$_{\alpha}$, broad Fe, and blackbody fluxes.
(b) Blackbody flux versus  Fe K$_{\alpha}$ and broad Fe fluxes.
Fluxes are in units of 10$^{-3}$ photons keV$^{-1}$ cm$^{-2}$ s$^{-1}$.}
\end{figure}

\begin{figure}
\plotone{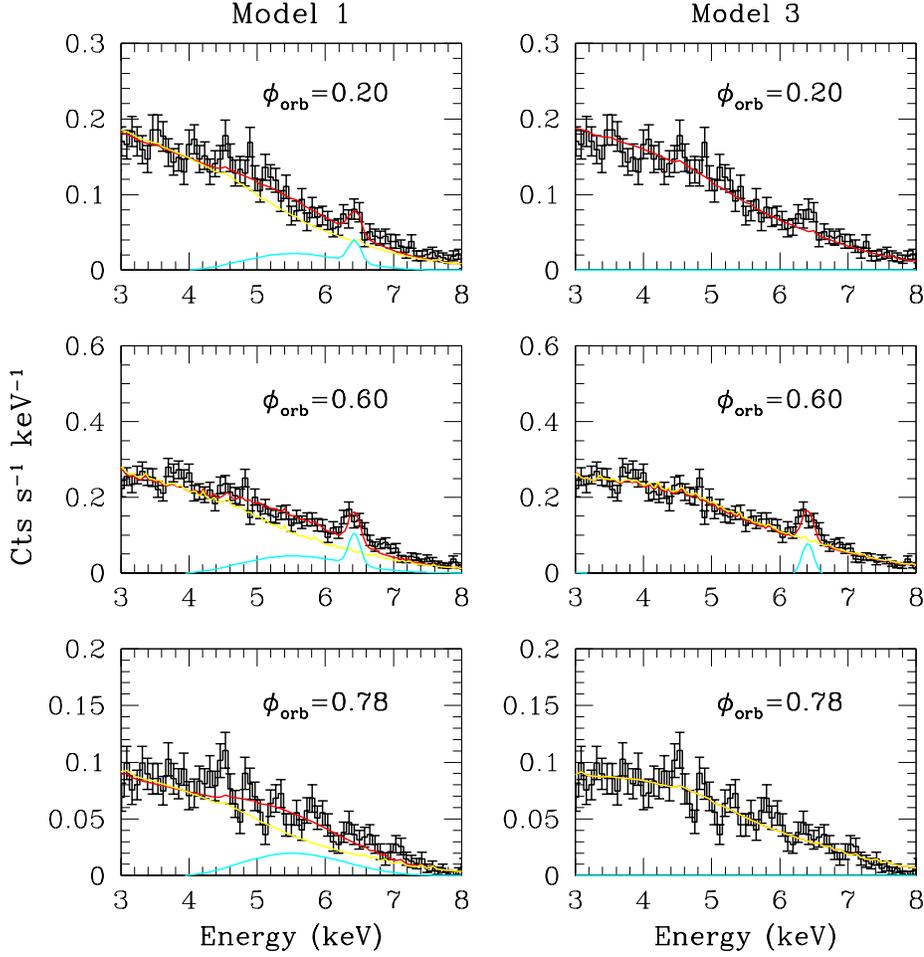}
\caption{
Left hand panels: Close-ups of the 3-8 keV segment of Figures 2-4 with the
the best fit Model 1 in red. The model for the Fe lines is that used by
Miller \etal~(2002) for Cyg X-1:  A narrow feature at 6.4 keV and
a broad ($\sigma$ = 800 eV) feature at 5.8 keV.  The blue line
is the fit with the Fe components excluded.  The magenta
shows the difference which comprises the Fe line components.
Right hand panels:  Close-ups of the 3-8 keV segment of Figures 2-4 with 
the best fit Model 3 in red.
}
\end{figure}

\begin{figure}
\plotone{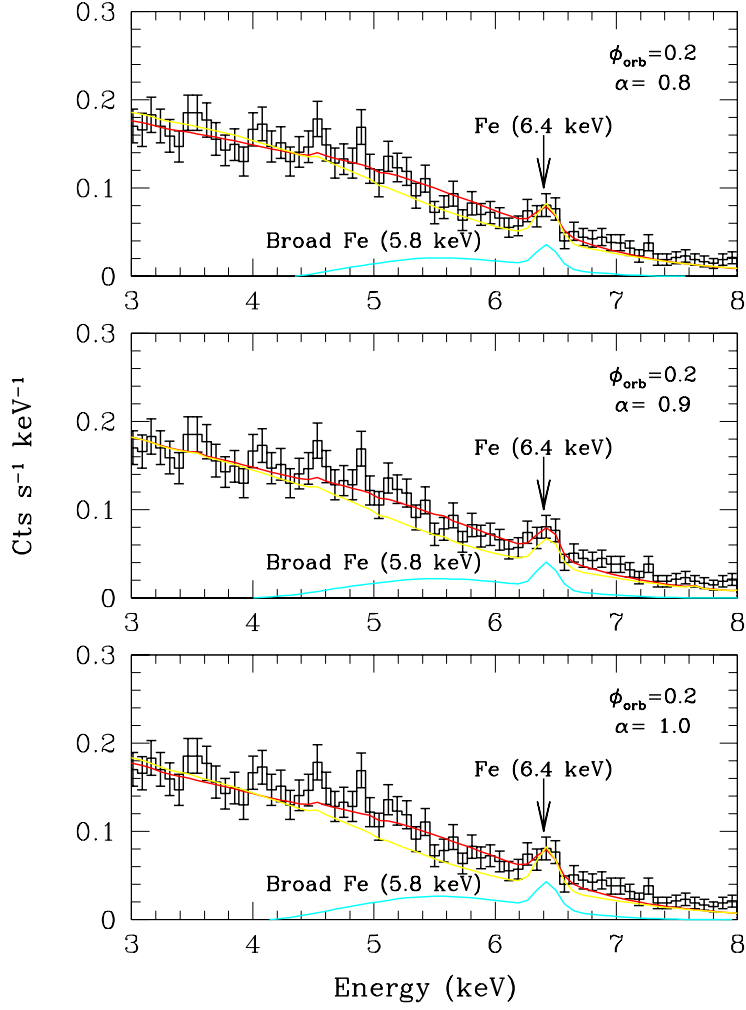}
\caption{
Close-ups of the 3-8 keV segment of Figure 2 with the
the best fit Model 1 in red. (a) For photon index $\alpha$ = 0.8.
(b) For photon index $\alpha$ = 0.9. 
(c) For photon index $\alpha$ = 1.0. 
1The model for the Fe lines is that used by
Miller \etal~(2002) for Cyg X-1:  A narrow feature at 6.4 keV and
a broad ($\sigma$ = 800 eV) feature at 5.8 keV.  The blue line
is the fit with the Fe components excluded.  The magenta
shows the difference which comprises the Fe line components.
}
\end{figure}

\begin{figure}
\plotone{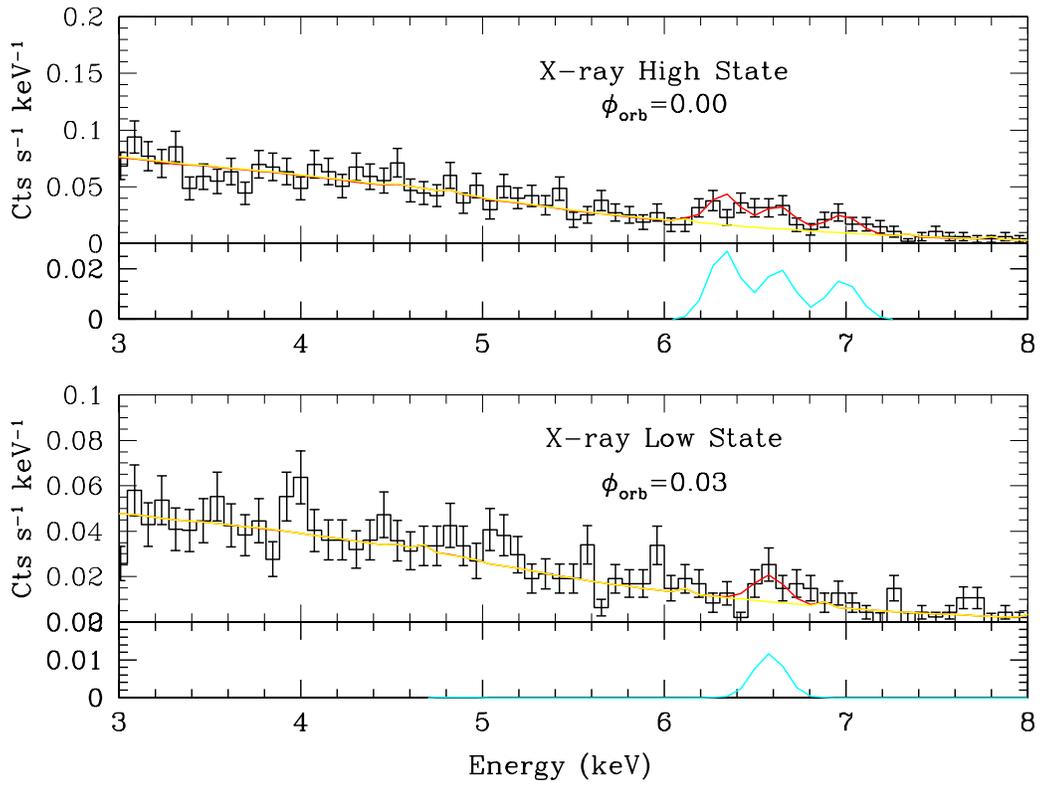}
\caption{
A close-up of the 3-8 keV segment of Figure 3 with the
the best fit model in red.
}
\end{figure}

\begin{figure}
\plotone{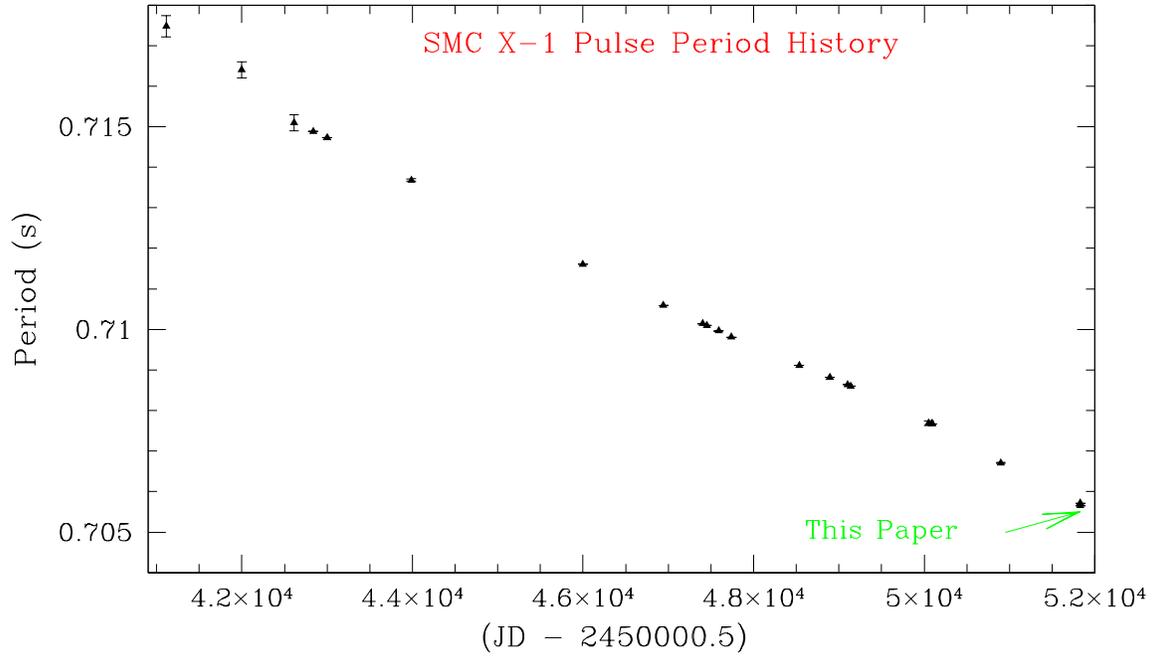}
\caption{
The best-fit for the pulse period (0.7057147$\pm$0.00000027s) shows that 
the spin-up trend of SMC X-1 continues.
}
\end{figure}






\clearpage


\begin{center}
\begin{tabular}{lcccc} \\
\multicolumn{5}{c}{Table 1: Log of SMC X-1 Chandra/ACIS Observations}\\
\hline \hline
Sequence&Julian Date&Good time&Orbital phase&X-ray\\
Number&at Start&(s)&at midpoint&State\\
\hline \hline
400101&2451831.3904&6194&0.00&High\\
400102&2451832.1694&6244&0.20&High\\
400103&2451833.3365&6133&0.50&High\\
400104&2451834.3083&6465&0.75&High\\
&&&&\\
400105&2451855.5201&6137&0.20&Low\\
400106&2451857.0972&6329&0.60&Low\\
&&&&\\
400107&2452025.1074&6124&0.78&Low\\
400108&2452026.0973&6156&0.03&Low\\
\hline \hline
\end{tabular}
\end{center}

\clearpage

\begin{center}
\begin{tabular}{lccc}
\multicolumn{4}{c}{Table 2a. Fits to High State Spectra: Model 1 (Power-law + BB)}\\
\hline \hline
Parameter&$\phi_{orb}$=0.20&$\phi_{orb}$=0.50&$\phi_{orb}$=0.75\\
\hline \hline
N$_H$ (10$^{22}$ cm$^{-2}$)&0.18$^{+0.01}_{-0.01}$&0.24$^{+0.01}_{-0.01}$&0.19$^{+0.01
}_{-0.01}$\\
I$_{pl}^{**}$& .056$^{+0.001}_{-0.001}$&0.058$^{+0.001}_
{-0.001}$&.056$^{+0.001}_{-0.001}$\\
$\alpha^*$&0.9&0.9&0.9\\
I$_{bb}^{**}$&.0012$^{+0.0001}_{-0.0001}$&.0019$^{+0.00
01}_{-0.0001}$&.0021$^{+0.0001}_{-0.0001}$\\
E$_{bb}$(keV)&0.18$^{+0.01}_{-0.01}$&0.19$^{+0.01}_{-0.01}$&0.18$^{+0.01}_{-0.01}$
\\
$\chi_{\nu}^2$&1.0&1.1&1.0\\
Pileup Fraction&0.16&0.17&0.16\\
\hline
\hline
\end{tabular}
\end{center}
$^*$Powerlaw photon index held at value found for SMC X-1 using Ginga
(Woo~\etal~1995) and ASCA
(Wojdowski~\etal~2000) observations.
$^{**}$(photons keV$^{-1}$cm$^{-2}$s$^{-1})$

\clearpage
\begin{center}
\begin{tabular}{lccc}
\multicolumn{4}{c}
{Table 2b. Fits to High State Spectra: Model 2(Cutoff Power-law + BB)}\\ 
\hline \hline
Parameter&$\phi_{orb}$=0.20&$\phi_{orb}$=0.60&$\phi_{orb}$=0.75\\
\hline \hline
N$_H$ (10$^{22}$ cm$^{-2}$)&0.16$^{+0.01}_{-0.01}$&0.25$^{+0.02}_{-0.02}$&0.16$^{+0.01
}_{-0.01}$\\
I$_{pl}$$^{**}$& .05$^{+0.01}_{-0.01}$&0.06$^{+0.01}_
{-0.01}$&.05$^{+0.01}_{-0.01}$\\
$\alpha^*$&1.0&1.0$^{+0.05}_{-0.03}$&0.3\\
E$_c$&6&32870$^{+60000}_{-32870}$&3\\
E$_f$&6&50$^{+4925}_{-49.5}$&6\\
I$_{bb}$$^{**}$&001$^{+0.0001}_{-0.0001}$&.0020$^{+0.00
01}_{-0.0001}$&.0011$^{+0.0002}_{-0.0001}$\\
E$_{bb}$(keV)&0.19$^{+0.002}_{-0.002}$&0.18$^{+0.01}_{-0.01}$&0.20$^{+0.01}_{-0.01}$
\\
$\chi_{\nu}^2$&1.0&1.1&1.0\\
Pileup Fraction&0.18&0.18&0.16\\
\hline
\hline
\end{tabular}
\end{center}
$^*$Powerlaw photon index held at value found for SMC X-1 using Ginga
(Woo~\etal~1995) and ASCA
(Wojdowski~\etal~2000) observations.
$^{**}$(photons keV$^{-1}$cm$^{-2}$s$^{-1})$
E$_c$ is the cutoff energy and E$_f$ the e-folding energy as defined by
(Woo~\etal~1995).

\clearpage
\begin{center}
\begin{tabular}{lccc}
\multicolumn{4}{c}{Table 2c. Fits to High State Spectra:
Model 3 (ComptT + BB)}\\
\hline \hline
Parameter&$\phi_{orb}$=0.20&$\phi_{orb}$=0.60&$\phi_{orb}$=0.75\\
\hline \hline
N$_H$ (10$^{22}$ cm$^{-2}$)&0.17$^{+0.01}_{-0.01}$&0.24$^{+0.01}_{-0.01}$&0.17$^{+0.01
}_{-0.01}$\\
I$_{ComptT}$& .045$^{+0.001}_{-0.001}$&0.008$^{+0001}_
{-0.01}$&.046$^{+0.001}_{-0.001}$\\
ComptT kT&110$^{+7}_{-30}$&496$^{+496}_{-496}$&102$^{+6}_{-25}$\\
ComptT TO&tied to E$_{bb}$&tied to E$_{bb}$&tied to E$_{bb}$\\
ComptT Tau&21.4$^{+0.05}_{-0.05}$&44$^{+1}_{-8}$&21.6$^{+0.05}_{-0.05}$\\
ComptT Redshift&fixed at 0&fixed at 0&fixed at 0\\
ComptT Approx&fixed at 2&fixed at 2&fixed at 2\\
I$_{bb}$&.001$^{+0.0001}_{-0.0001}$&.0019$^{+0.0002}_{-0.0002}$&.0021$^{+0.0001}_{-0.0
001}$\\
E$_{bb}$(keV)&0.16$^{+0.002}_{-0.002}$&0.18$^{+0.002}_{-0.003}$&0.18$^{+0.01}_{-0.01}$
\\
$\chi_{\nu}^2$&1.0&1.1&1.0\\
Pileup Fraction&0.18&0.18&0.18\\
\hline
\hline
\end{tabular}
\end{center}
$^{**}$(photons keV$^{-1}$cm$^{-2}$s$^{-1})$

\clearpage
\begin{center}
\begin{tabular}{llc} \\
\multicolumn{3}{c}{Table 3$^*$: Lines used in fitting 
low-state and eclipse spectra}\\
\hline \hline
Ion&Wavelength (\AA)&Energy (keV)\\
\hline\hline
O VII He$\alpha$&21.80$\star\bullet$&0.56\\
O VIII Ly$_{\alpha}$&18.97$\star\bullet$&0.65\\
O VIII Ly$_{\beta}$; O VII RRC& 16.5& 0.75 \\
Ne IX; OVIII RRC?&13.45$\star\bullet$&0.91\\
Ne X Ly$_{\alpha}$&12.13$\star\bullet$&1.02\\
Mg XI&9.17$\star\bullet\ast$&1.34\\
Mg XII Ly$_{\alpha}$&8.50$\star\bullet\ast\dagger$&1.47\\
Mg XII$Ly_{\beta}$&7.09$\ast\dagger$&1.75\\
Si XIII He$\alpha$ &6.70$\bullet\ast\dagger$&1.85\\
S XIV Ly$_{\alpha}$&6.20$\bullet\ast\dagger$&2.00\\
S XV; Si XIII RRC&5.20$\bullet\ast\dagger$&2.44\\
S XVI Ly$_{\alpha}$&4.75$\bullet\ast$&2.62\\
Si XIV RRC&4.65$\dagger$&2.66\\
Ar XVIII Ly$_{\alpha}$&3.75$\dagger$&3.30\\
Broad Fe&2.13&5.82\\
Fe K$_{\alpha}$&1.94$\bullet\ast\dagger$&6.38\\
Fe XXV He$\alpha$&1.86$\bullet\ast\dagger$&6.70\\
Fe XXVI Ly $_{\alpha}$&1.78$\bullet\ast\dagger$&7.00\\
\hline
\hline
\end{tabular}
\end{center}
\small
$^*$Lines marked with a: $\star$ have been seen in
XMM/RGS observations of SMC X-1 during eclipse (Wojdowski
2003);  $\bullet$ have been detected
in ASCA SIS observations of Vela X-1 (Sako~\etal~1997);
$\ast$ have been seen in Chandra/HETGS observations of Cen X-3
(Wojdowski~\etal~2003);
and with $\dagger$ have been
seen in Chandra/HETGS observations of Cyg X-3 (Paerels~\etal~2000).
\normalsize

\clearpage
\begin{center}
\begin{tabular}{lccc}
\multicolumn{4}{c}{Table 4a$^*$. Fits to Low State Spectra: (Model 1+L and Model 1+L+B
)}\\
\hline \hline
Orbital Phase&0.20&0.60&0.78\\
\hline\hline
N$_H$ (10$^{22})$cm$^{-2}$&0.30$^{+0.03}_{-0.03}$&0.42$^{+0.02}_{-0.05}$&0.29$^{+0.05}
_{-0.05}$\\
I$_{pl}$(photons keV$^{-1}$cm$^{-2}$s$^{-1})$&1.4$^{+0.4}_{-0.3}$e-3&2.1$^{+0.3}_{-0.4
}$e-3&6.9$^{+
0.2}_{-0.2}$e-4\\
$\alpha^{**}$&0.9&0.9&0.9\\
I$_{bb}$(photons keV$^{-1}$cm$^{-2}$s$^{-1})$&8.0$^{+1.9}_{-1.5}$e-5&3.9$^{+0.7}_{-0.7
}$e-4&3.2$^{+
1.0}_{-1.0}$e-5\\
E$_{BB}$(keV)&0.16$^{+0.005}_{-0.005}$&0.14$^{+0.003}_{-0.003}$&0.15$^{+0.01}_{-0.01}$\\
\hline
O VII  0.56&...&1.4$^{+0.2}_{-0.3}$e-3&...\\
O VIII  0.65 &4.3$^{+1.4}_{-1.4}$e-4&1.3$^{+0.4}_{-0.4}$e-3&3.1$^{+0.1}_{-0.1}$e-4\\
Ne IX   0.91 &2.8$^{+0.4}_{-0.5}$e-4&2.0$^{+0.1}_{-0.1}$e-3&1.5$^{+0.3}_{-0.3}$e-4\\
Mg XI  1.34&...&2.6$^{+0.2}_{-0.3}$e-4&...\\
Mg XII  1.84 &2.7$^{+1.2}_{-0.9}$e-5&9.8$^{+0.7}_{-1.3}$e-5&...  \\
Si XIII RRC 2.44 &3.5$^{+1.4}_{-1.0}$e-5&5.9$^{+1.5}_{-1.4}$e-5&...\\
Broad Fe  5.8 & 2.1$^{+0.5}_{-0.3}$e-4 &4.3$^{+0.6}_{-0.4}$e-4&1.9$^{+0.4}_{-0.2}$e-4\\
Fe K$_\alpha$ 6.38   &4.4$^{+1.5}_{-1.5}$e-5&1.2$^{+0.2}_{-0.2}$e-4&...\\
\hline
$\chi_{\nu}^2$ without broad Fe       &1.7   &1.6   &1.4\\
$\chi_{\nu}^2$ with broad Fe   &1.2   &1.2   &1.0 \\
Pileup fraction         &.007  &.014  &.003\\
\hline \hline
\end{tabular}
\end{center}
$^*$All line widths were held constant at 50 eV.  The broad Fe feature
was held at a width of 800eV.
Errors listed are 1$\sigma$. Lines detected below 3$\sigma$ are not listed.\\
$^{**}$Powerlaw photon index held at value found for SMC X-1 using Ginga
(Woo~\etal~1995), ASCA
(Wojdowski~\etal~2000), and BeppoSAX (REF) observations.

\clearpage
\begin{center}
\begin{tabular}{lccc}
\multicolumn{4}{c}{Table 4b$^*$. As in 4a for $\phi_{orb}$ = 0.2 for extremes
of Powerlaw photon index
)}\\
\hline \hline
N$_H$ (10$^{22})$cm$^{-2}$&0.29$^{+0.03}_{-0.03}$&0.30$^{+0.03}_{-0.03}$&0.30$^{+0.02}_{-0.05}$\\
I$_{pl}$(photons keV$^{-1}$cm$^{-2}$s$^{-1})$&1.2$^{+0.03}_{-0.04}$e-3&1.4$^{+0.03}_{-0.04}$e-3&1.5$^{+0.4}_{-0.3}$e-3\\
$\alpha^{**}$&0.8&0.9&1.0\\
I$_{bb}$(photons keV$^{-1}$cm$^{-2}$s$^{-1})$&8.0$^{+1.9}_{-1.5}$e-5&8.0$^{+2.0}_{-1.5}$e-5&8.0$^{+0.7}_{-0.7
}$e-5\\
E$_{BB}$(keV)&0.17$^{+0.005}_{-0.005}$&0.16$^{+0.005}_{-0.005}$&0.16$^{+0.003}_{-0.003}$\\
\hline
O VIII  0.65 &4.1$^{+1.2}_{-1.4}$e-4&4.3$^{+1.4}_{-1.4}$e-4&4.2$^{+1.6}_{-0.9}$e-4\\
Ne IX   0.91 &2.6$^{+0.4}_{-0.5}$e-4&2.8$^{+0.4}_{-0.5}$e-4&2.8$^{+0.5}_{-0.4}$e-4\\
Mg XII  1.84 &3.6$^{+1.1}_{-0.8}$e-4&2.7$^{+1.2}_{-0.9}$e-5&2.8$^{+1.0}_{-0.8}$e-5\\
Si XIII RRC 2.44 &3.9$^{+1.3}_{-1.0}$e-5&3.5$^{+1.4}_{-1.0}$e-5&3.7$^{+1.3}_{-0.9}$e-5\\
Broad Fe  5.8 & 1.5$^{+0.5}_{-0.2}$e-4&2.1$^{+0.5}_{-0.2}$e-4 &2.6$^{+0.4}_{-0.2}$e-4\\
Fe K$_\alpha$ 6.38   &4.4$^{+1.4}_{-1.3}$e-5&4.4$^{+1.5}_{-1.5}$e-5&4.4$^{+1.6}_{-1.2}$e-5\\
\hline
$\chi_{\nu}^2$ without broad Fe &1.5      &1.7   &2.1   \\
$\chi_{\nu}^2$ with broad Fe &1.3  &1.2   &1.5    \\
Pileup fraction         &.007  &.007  &.007\\
\hline \hline
\end{tabular}
\end{center}
$^*$All line widths were held constant at 50 eV with the exception of the
broad Fe feature which was held at 800 eV. 
Errors listed are 1$\sigma$. Lines detected below 3$\sigma$ are not listed.

\clearpage
\centerline{Table 4c$^*$. Fits to Low State data: Model 3 (ComptT)}
\begin{center}
\begin{tabular}{lccc}
\hline \hline
Model &0.20&0.60&0.78\\
\hline\hline
N$_H$ (10$^{22})$cm$^{-2}$&0.40$^{+0.02}_{-0.05}$&0.33$^{+0.02}_{-0.04}$&0.22$^{+0.04}
_{-0.06}$\\
I$_{comptt}^{**}$&2.99$^{+0.5}_{-0.6}$e-3&3.0$^{+0.4}_{-0.2}$e-3&1.0$^{+0.1}_{-0.3}$e-
3\\
CompttkT&6$^{+5}_{-1}$&6.9$^{+2.2}_{-1.9}$&4.3$^{+1.9}_{-1.0}$\\
CompttTO&tied to E$_{BB}$&tied to E$_{BB}$&tied to E$_{BB}$\\
ComptTau&23$^{+120}_{-2}$&28$^{+3}_{-2}$&30$^{+6}_{-2}$\\
ComptRedshift&held at 0&held at 0&held at 0\\
ComptApprox&held at 2&held at 2&held at 2\\
I$_{bb}^{**}$&2.2$^{+0.7}_{-0.7}$e-4&2.1$^{+
0.9}_{-0.4}$e-4&1.8$^{+0.8}_{-0.5}$e-5\\
E$_{BB}$(keV)&0.14$^{+0.003}_{-0.003}$&0.16$^{+0.03}_{-0.01}$e-3&0.19$^{+0.002}_{-0.01
}$\\
\hline
O VII  0.56&...&6.3$^{+0.4}_{-1.2}$e-3&...\\
O VIII  0.65 &...&2.3$^{+0.7}_{-0.7}$e-3&3.0$^{+0.3}_{-0.3}$e-4\\
Ne IX   0.91 &...&1.8$^{+0.1}_{-0.1}$e-3&1.4$^{+0.3}_{-0.3}$e-4\\
Mg XI  1.34&...&2.0$^{+0.3}_{-0.1}$e-4&...\\
Mg XII  1.84 &...&1.1$^{+0.1}_{-1.5}$e-4&...  \\
Si XIII  2.40 &...&8.2$^{+1.1}_{-1.6}$e-5&...\\
Broad Fe  5.8 & ...&...&...\\
Fe K$_\alpha$ 6.38   &...&1.2$^{+0.5}_{-0.4}$e-4&...\\
\hline
$\chi_{\nu}^2$  &1.2   &1.1   &1.0 \\
Pileup fraction    &0.0&0.0  &0.0  \\
\hline \hline
\end{tabular}
\end{center}
\small
$^*$All line widths were held constant at 50 eV.
Errors listed are 1$\sigma$. Lines detected below 3$\sigma$ are not listed.\\
$^{**}$ (photons keV$^{-1}$cm$^{-2}$s$^{-1}$)

\clearpage
\begin{center}
\begin{tabular}{lcc}
\multicolumn{3}{c}{Table 5$^*$: Fits to Eclipse Spectra}\\
\hline \hline
Orbital Phase&0.00&0.03\\
\hline\hline
N$_H$ (10$^{22})$cm$^{-2}$&0.21$^{+0.03}_{-0.03}$&0.57$^{+0.06}_{-0.09}$\\
I$_{pl}$(photons keV$^{-1}$cm$^{-2}$s$^{-1})$&5.8$^{+0.2}_{-0.2}$e-4&3.9$^{+0.1}_{-0.2
}$e-4\\
$\alpha^{**}$&0.94&0.94\\
I$_{bb}$(photons keV$^{-1}$cm$^{-2}$s$^{-1})$&1.3$^{+0.6}_{-0.3}$e-5&1.2$^{+0.5}_{-0.5
}$e-4\\
E$_{BB}$(keV)&0.22$^{+0.04}_{-0.01}$&0.11$^{+0.01}_{-0.01}$\\
\hline
O VIII  0.65 &4.5$^{+1.0}_{-0.8}$e-4&...\\
Ne IX   0.91 &2.1$^{+0.3}_{-0.3}$e-4&2.1$^{+0.7}_{-0.5}$e-4\\
Ne X    1.02&9.0$^{+1.9}_{-2.0}$e-5&...\\
Mg XI  1.34&4.5$^{+1.2}_{-1.1}$e-5&3.6$^{+1.3}_{-1.3}$e-5\\
Fe K$_\alpha$ 6.38   &2.4$^{+0.9}_{-0.9}$e-5&...\\
Fe XXV 6.6&3.4$^{+1.1}_{-1.1}$e-5&1.9$^{+0.4}_{-0.5}$e-5\\
Fe LY$_{\alpha}$ 7.0&3.6$^{+1.3}_{-1.3}$e-5&...\\
\hline
$\chi_{\nu}^2$ &0.9    &1.0 \\
Pileup fraction          &.000  &.001\\
\hline \hline
\end{tabular}
\end{center}
$^*$All line widths were held constant at 50 eV.
Errors listed are 1$\sigma$. Lines detected below 3$\sigma$ are not listed.
$^{**}$Powerlaw photon index held at value found for SMC X-1 using Ginga
(Woo~\etal~1995) and ASCA
(Wojdowski~\etal~2000) observations.







\end{document}